\documentclass[preprint2]{aastex}

\newcommand{\nn}{[\mbox{N\,{\sc ii}}]}
\newcommand{\ltsim}{\lower.5ex\hbox{$\; \buildrel < \over \sim \;$}}
\newcommand{\gtsim}{\lower.5ex\hbox{$\; \buildrel > \over \sim \;$}}

\newcommand{\ha}{H$\rm\alpha$}
\newcommand{\kms}{~km~s$^{-1}$}
\newcommand{\nb}{[\mbox{N\,{\sc ii}}]~$\lambda$6583}
\newcommand{\nc}{[\mbox{N\,{\sc ii}}]~$\lambda$6548}
\begin{document}

\title{Is the Broad-Line Region Clumped or Smooth? \\
Constraints from the \ha\ Profile in NGC 4395, the Least Luminous
Seyfert 1 Galaxy}

\author{Ari Laor\altaffilmark{1}, Aaron J. Barth\altaffilmark{2},
Luis C. Ho\altaffilmark{3}, and Alexei V. Filippenko\altaffilmark{4}}
\altaffiltext{1}{Physics Department, Technion, Haifa 32000, Israel;
laor@physics.technion.ac.il}
\altaffiltext{2}{Department of Physics \& Astronomy
4129 Frederick Reines Hall
University of California,
Irvine, CA 92697-4575; barth@uci.edu}
\altaffiltext{3}{The Observatories of the Carnegie Institution of Washington,
813 Santa Barbara Street, Pasadena, CA 91102; lho@ociw.edu}
\altaffiltext{4}{Department of Astronomy, 601 Campbell Hall, 
University of California, Berkeley, CA 94720-3411; alex@astro.berkeley.edu}

\begin{abstract}
The origin and configuration of the gas which emits broad lines in
Type I active galactic nuclei is not established yet. The lack of
small-scale structure in the broad emission-line profiles is consistent
with a smooth gas flow, or a clumped flow with many
small clouds. An attractive possibility for the origin of many small clouds
is the atmospheres of bloated stars, an origin which also provides a natural
mechanism for the cloud confinement. Earlier studies of the broad-line profiles
have already put strong
lower limits on the minimum number of such stars, but these limits are sensitive
to the assumed width of the lines produced by each cloud. Here we revisit
this problem using high-resolution Keck spectra of the \ha\ line in NGC~4395,
which has the smallest known broad-line region ($\sim 10^{14}$~cm). Only a
handful of the required bloated stars (each having $r_{*} \approx 10^{14}$~cm)
could fit into the broad-line region of NGC~4395, yet the observed smoothness of
the \ha\ line implies a lower limit of
$\sim 10^4-10^5$ on the number of discrete clouds. This rules out conclusively the
bloated-stars scenario, regardless of any plausible line-broadening mechanisms.
The upper limit on the size of the clouds is $\sim 10^{12}$~cm,
which is comparable to the size implied by photoionization models.
This strongly suggests that gas in the broad-line region is
structured as a smooth rather than a clumped flow, most likely in a rotationally
dominated thick disk-like configuration. However, it remains to be clarified 
why such a smooth, gravity-dominated flow generates double-peaked emission 
lines only in a small fraction of active galactic nuclei.

\end{abstract}

\keywords{galaxies: active --- galaxies: Seyfert --- quasars: emission lines
--- galaxies: individual \objectname{NGC 4395}}

\section{Introduction}
Type I active galactic nuclei (AGNs) are defined by the presence of permitted lines
with a kinematically distinct broad component, usually with full width at 
half-maximum intensity (FWHM) $\gtsim 1000$~km~s$^{-1}$. The broad lines are 
emitted by gas in the broad-line region (BLR) which is being photoionized
by the central compact continuum source. Detailed
photoionization calculations yield rather tight constraints on the
density, temperature, ionization level, and chemical abundances of the
photoionized gas (e.g., Ferland et al. 1998; Hamann \& Ferland 1999), while
reverberation mapping techniques constrain the size and velocity field
of the gas (e.g., Peterson 2001). However, the structure and origin of the
photoionized gas in the BLR remains a largely unsolved problem.

\subsection{Proposed Models for the BLR}
The gas in the BLR may be structured in small clumps
(hereafter ``clouds''). A particularly appealing model is that
of bloated stars, where the clouds are identified with the surface layers
of supergiant stars, which are further bloated by the ionizing radiation field
(Edwards 1980; Scoville \& Norman 1988; Penston 1988; Kazanas 1989;
Tout et al. 1989; Alexander \& Netzer 1994, 1997). This model is attractive
because it relies on a relatively small modification of known objects (supergiants), and it
also provides a natural mechanism for the cloud confinement, although a self-consistent
BLR structure within this model appears problematic (Begelman \& Sikora 1992; Krolik 1999).

Alternative suggestions for the origin of BLR clouds include
gas streams produced by tidally disrupted stars (Hills 1975; Roos 1992),
or by star-disk collisions (Zurek et al. 1994; Vilkoviskij \& Czerny 2002),
or clumps in a gravitationally unstable outer
disk (Collin \& Hur{\' e} 2001). Other scenarios invoke density
inhomogeneities produced by wind interaction with nearby obstacles
(Perry \& Dyson 1985), such as stellar atmospheres
(Torricelli-Ciamponi \& Pietrini 2002), supernova remnants
(Pittard et al. 2003), or the the accretion-disk surface (Cassidy \& Raine 1996).
Alternatively, it was suggested that the clouds are density enhancements produced by
shocks in an accreting gas (Fromerth \& Melia 2001). The physical
viability of some of the proposed mechanisms remains unclear (e.g., Mathews 1986;
Mathews \& Doane 1990)

The other option for the BLR structure is a smooth continuous flow, rather
than a clumped flow. For example, this flow may be identified
with the illuminated and photoionized accretion-disk surface
(e.g., Dumont \& Collin-Souffrin 1990a,b). Alternatively, it could originate
in a wind driven off an accretion disk by
radiation or magnetic pressure gradients (Shlosman et al. 1985;
Emmering et al. 1992; K{\" o}nigl \& Kartje 1994; Murray et al. 1995;
Chiang \& Murray 1996; Bottorff et al. 1997).

\subsection{Earlier Constraints}

The minimum size of the photoionized clouds can be estimated from
simple Str\"{o}mgren depth arguments. Equating the H ionization rate to the
recombination rate per unit area yields the column density of the photoionized
H~II layer, where most of the line emission is produced. This argument yields
$\Sigma_{\rm ion} \approx 10^{23}U$~cm$^{-2}$, where $U\equiv n_\gamma/n_e$
is the ionization parameter, and $n_\gamma$ and $n_e$ are (respectively) the
ionizing photon and the electron number densities. Typical BLR parameters are 
$U \approx 0.1$ and $n_e \approx 10^{10}$~cm$^{-3}$, implying a typical thickness of 
the H~II layer of $d \approx 10^{12}$~cm (although there may be a large range of 
$d$ values in the BLR; e.g., Baldwin et al. 1995).
The equivalent widths of the H recombination lines
imply that the BLR gas covers $\Omega_{\rm BLR}/4\pi \approx 0.1-0.2$ of the ionizing
continuum source (e.g., Ferland et al. 1998). The photoionized gas could
be made of clumps which are as small as $d$, or it may reside in
a thin photoionized surface layer over more extended bodies. The latter
option is supported by the strength of some of the low-ionization lines (e.g.,
lines from Fe~II, Ca~II, Na~I) which
originate in deeper layers at $\Sigma> \Sigma_{\rm ion}$, where H is only
partially ionized (e.g., Ferland \& Persson 1989; Baldwin et al. 2003).
{\em How can one then determine if the BLR is clumped or smooth?}

The number of clouds in the BLR, $n_c$, and their size, $r_c$, can be
constrained based on the smoothness of the emission-line profiles. The
level of fluctuations in the emission-line profiles is proportional to
$1/\sqrt{n_c}$, assuming the cloud velocities are randomly distributed
within the line profile (in proportion to the line flux). The solid area
subtended by all the BLR clouds is
$\Omega_{\rm BLR}/4\pi=n_c\pi r_c^2/4\pi R_{\rm BLR}^2$, where $R_{\rm BLR}$
is the radius of the BLR; since $\Omega_{\rm BLR}/4\pi \approx 0.1-0.2$, we find that
$r_c\simeq R_{\rm BLR}/\sqrt{n_c}$. High-resolution spectroscopy at a high 
signal-to-noise ratio (S/N)
can place tight constraints on the level of fluctuations, which then translates
into a large lower limit on $n_c$, and thus a tight upper limit on $r_c$, which may
exclude some clumped BLR models.

The first application of this method  was made by
Capriotti et al. (1981), who deduced $n_c\gtsim 10^4$ in a sample of
Seyfert galaxies. Atwood et al. (1982) improved the limit to
$n_c\gtsim 5\times 10^4$ for Mrk~509, and Laor et al. (1994)
obtained $n_c\gtsim 10^4$ for a sample of bright quasars, using
{\em Hubble Space Telescope}  spectroscopy of the C~IV and
Ly~$\alpha$ lines. Significantly
improved limits of $n_c \gtsim 3\times 10^6$ and $3\times 10^7$
were obtained for Mrk~335 (Arav et al. 1997) and NGC~4151 (Arav et al. 1998)
using high-quality Keck spectra of the
\ha\ emission line. Finally, Dietrich et al. (1999) obtained the highest
limit of $n_c\gtsim 10^8$ in the luminous quasar 3C~273.

Since  $R_{\rm BLR}\propto L_{\rm opt}^{\sim 0.6}$ (Kaspi et al. 2005),
while $r_c \approx R_{\rm BLR}/\sqrt{n_c}$, tighter constraints on
$r_c$ can be obtained in lower-luminosity objects. In 3C~273,
$\nu L_{\nu}(5100$\AA$) \approx 6\times 10^{45}$~erg~s$^{-1}$,
and the lowest luminosity object where $n_c$ was constrained is NGC~4151,
where $\nu L_{\nu}(5100$\AA$)=7\times 10^{42}$~erg~s$^{-1}$ (Kaspi et al. 2000).

Here we report on high-quality Keck spectroscopy of the \ha\ line
in NGC~4395, which is a factor of 1000 fainter than NGC~4151, having
$\nu L_{\nu}(5100$\AA$)=5.9\times 10^{39}$~erg~s$^{-1}$ (Peterson et al. 2005).
This is the least luminous known Seyfert 1 nucleus (Filippenko \& Sargent 1989;
Filippenko, Ho, \& Sargent 1993), and despite its extreme low luminosity, it
appears to have a rather normal broad emission-line spectrum (apparently with a
lower than solar metal abundance, Kraemer et al. 1999).
The low luminosity of NGC~4395 implies an expected $R_{\rm BLR}$ of just $\sim
(0.5-1.7)\times 10^{14}$~cm, based on the most recent expression for
the mean Balmer-line time lag versus $\nu L_{\nu}(5100$\AA) in Kaspi et al. (2005;
see their Table 3). Bloated stars are predicted to have a size of
$\sim 10^{14}$~cm (Alexander \& Netzer 1994), and thus at most a handful
of such stars can be accommodated in the BLR of NGC~4395. This will
inevitably produce large-amplitude features in the line profile. 

The earlier studies of Arav et al. (1997, 1998) and Dietrich et al. (1999) ruled
out the bloated-stars idea, assuming the width of the lines produced by the
individual clouds are not much above the thermal width.
However, significant nonthermal broadening through radiation pressure
ablation of the BLR clouds, or through magnetic pressure effects,
is likely and possibly inevitable. This broadening reduces the lower limit on
$n_c$ (e.g., Bottorff \& Ferland 2000;
Bottorff et al. 2000), and it may leave the bloated-stars idea as a viable option.
As we show
below, the bloated-stars idea can be conclusively ruled out in the compact BLR of
NGC~4395, regardless of the velocity width of the lines
produced by the individual clouds. The upper limit obtained is comparable to $d$,
strongly suggesting that the broad
lines are produced by a smooth gas flow. In \S 2 of this paper we describe the
observations and method of analysis, and the results are discussed in
\S 3.

\section{Observations and Analysis}

The spectra we use here were obtained in three observations of NGC~4395
with the Keck-I and Keck-II 10~m telescopes. Two of the observations were made
with the
High Resolution Echelle Spectrometer (HIRES; Vogt et al. 1994); the spectral 
resolution was $\lambda/\Delta\lambda = 38,000$, or 0.173~\AA\ near \ha, 
sampled at 0.0473~\AA\ per pixel. The third observation was made with the 
Echellette Spectrograph and Imager (ESI; Sheinis et al. 2002); the spectral 
resolution was $\lambda/\Delta\lambda=8,000$, or 0.820~\AA\ near \ha, 
sampled at 0.259~\AA\ per pixel. The first HIRES observation (hereafter 
HIRES1), described by Filippenko \& Ho (2003), was a 20 + 40 minute 
exposure obtained on 1994 April 14 UT, which yielded a S/N ranging from 
20 per pixel at the continuum near \ha\ to 300 at the line peak. The 
second HIRES observation (hereafter HIRES2) was a 30 + 60 minute exposure 
taken on 2002 February 4 UT, yielding a S/N ranging from 30 to 300, as above.
The ESI observation was a 20 + 20 minute exposure made on 2002 December 2 UT,
and the S/N ranges from 50 to 400, as above.

All of the spectra were reduced using the MAKEE reduction software
\footnote{see www2.keck.hawaii.edu/inst/hires/makeewww/index.html}.
Cosmic-ray hits were manually removed from the spectra by comparing
the two exposures available for each of the spectra. We did not use
standard cosmic-ray removal tasks as they sometimes leave small residual features,
which will affect the fluctuation analysis. We
used a simple summation for the extractions rather than an optimal
extraction in order to preserve the correct emission-line profiles,
since an optimal extraction might produce incorrect results for the
narrow-line profiles, and it could potentially produce small distortions
in the broad lines as well.

The rest-frame wavelength scale was converted to a velocity scale with respect
to the rest wavelength of \ha, 6562.8~\AA. We applied
a slight offset to the velocity scale in each spectrum, such that the narrow \ha\ peak
falls at $v=0$\kms\ in all cases.
We next subtracted a constant continuum level from each spectrum, in order to
get the net line flux. Absolute flux calibration is available only for the
ESI spectrum, and an arbitrary flux scale is used for the two HIRES spectra.
The lack of an absolute flux calibration does not affect our analysis
as we are only interested in the \ha\ profile shape here.

The minimum number of clouds is estimated as follows. We first find for each
observed profile, $f_{\lambda}^{\rm obs}$, the smoothest acceptable fit,
$f_{\lambda}^{\rm fit}$ --- that is, a fit which yields $\chi^2_r \approx 1$ for the
given flux measurement errors, $f_{\lambda}^{\rm err}$. We next produce a Monte Carlo (MC)
realization of the observed profile by adding the emission of $n_c$ lines,
each one having a Gaussian line shape with a velocity dispersion $\sigma$.
Each line is further convolved with the instrumental resolution, approximated as
a Gaussian with FWHM = 8~\kms\ for HIRES, and 38~\kms\ for ESI.
We then measure $\chi^2_r$ of the MC simulated profile, $f_{\lambda}^{\rm sim}$,
with respect to the smooth-fit profile, as a function of $n_c$ and $\sigma$.
For a given $\sigma$, we find the minimum $n_c$ required to obtain
$\chi^2_r \approx 1$. This procedure provides us with an estimate of the minimum
$n_c$, as a function of $\sigma$, required so that the level of
fluctuations produced by the finite number of clouds does not exceed the observed
level of fluctuations (produced by the flux measurement errors).

To obtain the smooth model fit, each line is fit with a spline function, where we
iterate  manually over the number and positions of the points selected for the spline
fit, until the fit yields a reduced $\chi^2_r \approx 1$
over the velocity range of $-$2500\kms\ to 2500\kms (or smaller, as set
by the available spectra). The error
spectrum, $f_{\lambda}^{\rm err}$, obtained together with
$f_{\lambda}^{\rm obs}$ in the MAKEE pipeline reduction, is used for calculating
the $\chi^2$ of the spline fit.
The broad \ha\ line is blended with the  narrow \ha\ component and
with the narrow \nc\ and \nb\ doublet lines. To make sure that the spline
fit is made only in regions dominated by the broad-component emission, we add
the additional constraint that the residual flux in the two narrow \nn\
line profiles should match.\footnote{The measured flux ratio of
\nb/\nc\ in the spectra is $\sim 3$, which agrees with the
theoretically predicted value} The \nn\ profile was also used as a guide
for the expected shape of the narrow \ha\ component, and the broad \ha\
spline fit was constructed to yield a narrow \ha\ profile similar (but
not identical) to the \nn\ profile. This procedure allowed us to determine
the velocity ranges  affected by narrow-line contamination, which we find are
 $-900$\kms$ < v < -550$ \kms, $-210$ \kms $ < v < 230$ \kms, and
 $700$ \kms $ < v < 1200$ \kms. We exclude these velocity ranges 
in the $\chi^2$ calculation.

Figure 1 displays the HIRES1, HIRES2, and ESI \ha\ profiles, 
the smooth fits, and the residuals
from the fit. Achieving $\chi^2_r \approx 1$ required about 14 spline
points per spectrum. These were spaced about $150$~\kms\ apart close to the
line core (between the \nn\ lines), and about $250$~\kms\ or more apart
in the line wings. The $\chi^2$ and the number of degrees of freedom obtained through
the spline fit are 432/400 for ESI, 1118/1446 for HIRES1, and 688/812 for HIRES2.
Thus, the fluctuations on scales smaller than the spline smoothing length are
consistent with the level expected from $f_{\lambda}^{\rm err}$. There is
no evidence for significant intrinsic profile structure beyond pure 
noise.\footnote{This also rules out the presence of detectable stellar absorption
features from the underlying host galaxy light.}
There are significant, though weak (amplitude $\sim 1$\%),
features in the broad-line profiles close to the core on scales $\gtsim 150$~\kms,
as indicated by small
and non-monotonic changes in the smooth-fit profile curvature.
At the wings the smooth fit appears featureless; however, since
the S/N drops, such broad weak features would not be detectable there.

For the MC simulation, each one of the $n_c$ clouds is assumed to emit a
line with a Gaussian profile with the same total flux and velocity dispersion
$\sigma$. The clouds differ only in their centroid velocity $v$.
We next need the velocity probability distribution function, $P(v)$, for the
MC simulation. This $P(v)$ should give
$f_{\lambda}^{\rm sim}\rightarrow f_{\lambda}^{\rm fit}$
in the limit of $n_c\rightarrow\infty$.
As an estimate for $P(v)$ we use the inverse function of
$$g_v=C\int_{-\infty}^{v}f_vdv$$
\noindent 
($C$ is set
such that $g_{\infty}=1$, as required for the integral of a distribution
function), where $f_v$ is $f_{\lambda}^{\rm fit}$ converted to a
velocity scale. The use of an inverse function is a general procedure allowing
one to transform a uniform random distribution from 0 to 1
(generated by any standard random-number generator) into any given normalized
distribution ($f_v$ here; see also Press et al. 1992, Chapter 7.2).
The MC profile, $f_{\lambda}^{\rm sim}$,
is generated by adding the contribution of the $n_c$ individual profiles, each one
at its respective $v$.
The goodness of fit for each MC realization is measured by calculating
the $\chi^2$ of $f_{\lambda}^{\rm sim}$, with respect to
$f_{\lambda}^{\rm fit}$, using $f_{\lambda}^{\rm err}$ for the flux errors.
Since each MC realization yields a somewhat different value of 
$\chi^2$, we performed 20 different MC simulations for
each set of $n_c$ and $\sigma$ values, and obtained  the average value
$\chi^2_{r,\rm av}$ and  $\Delta\chi^2_r$, the dispersion of $\chi^2_r$ in the 20
simulations, which is used for verifying that $\chi^2_{r,\rm av}$ is accurate enough
(i.e., $\Delta\chi^2_r/\chi^2_{r,\rm av}\ll 1$).
A model was defined as acceptable if $\chi^2_{r, \rm av}=1\pm \Delta\chi^2_r$.

The observed line profile is a convolution of the cloud velocity distribution
function, $P(v)$, and the line profile produced by each cloud.
Above we have made the approximation that $P(v)$ is proportional to the observed
profile, and the accuracy of this approximation drops as $\sigma$ increases.
Specifically, we find that for $b\ge 50$~\kms\ ($b=\sqrt{2}\sigma$)
the above procedure does
not yield $f_{\lambda}^{\rm sim}\rightarrow f_{\lambda}^{\rm fit}$ for
$n_c\rightarrow\infty$. The simulated line peak becomes broader than the observed
peak due to the effect of the convolution.
Another difficulty arises from the fact that the spline fit we use follows
profile features on scales $\gtsim 150$~\kms, and thus features on this
scales are filtered out of the observed residuals spectrum
($f_{\lambda}^{\rm obs}-f_{\lambda}^{\rm fit}$). On the other hand, a MC simulation of
clouds with
a given $\sigma$ produces spectral features on a scale of $\sigma$.
Thus, for large $\sigma$, $f_{\lambda}^{\rm sim}-f_{\lambda}^{\rm fit}$
shows large-scale features, while such features do not exist in the observed residuals
spectrum.
These features can produce significant $\chi^2$ even for large values of $n_c$. To overcome
these two problems we added the following iteration. We took $f_{\lambda}^{\rm sim}$
obtained for a specific simulation, and fit it with a spline
(using the same velocities used in the observed line fit), in order to get a revised
estimate of $f_{\lambda}^{\rm fit}$. We now use the revised $f_{\lambda}^{\rm fit}$,
to measure the $\chi^2_r$ of $f_{\lambda}^{\rm sim}$ with respect to the revised
$f_{\lambda}^{\rm fit}$.

Figure 2 demonstrates the effect of $n_c$ and $\sigma$ on the level of
fluctuations in the MC simulations, compared to the observed residuals
(shown for the HIRES2 spectrum). As expected, the simulated profile
becomes smoother with increasing $n_c$ and $\sigma$. In particular, for
$b=10$~\kms, which corresponds to pure thermal broadening of $T \approx 10^4$~K
gas, the MC fluctuations are comparable to the observed ones for $n_c=10^5$,
while for
$b=100$~\kms, $n_c=3\times 10^3$ is enough to produce fluctuations which
are about comparable
to the observed ones. A more quantitative measure of the fluctuations level
is obtained by calculating the $\chi^2$ of the fluctuations, as mentioned above.

Figure 3 presents our derived limits on the minimum $n_c$, as a function of $b$,
required to yield
$\chi^2_r\le 1$ for the HIRES1/2 spectra and for the ESI spectrum. The
strongest constraints are provided by the ESI spectrum, where the broad \ha\
line happens to have the largest strength relative to the narrow \ha\ line. For $b=10$~\kms\
the minimum number of clouds required is $1.35\times 10^5$; this drops to 
$2.4\times 10^4$ for $b=100$~\kms, and to 3900 for $b=200$~\kms.
The solid lines in the plot follow the
$n_c\propto b^{-1}$ relation, expected from simple fluctuation
arguments.\footnote{The number of clouds contributing to each velocity bin
in the spectrum is $\propto n_c$ and also $\propto b$, i.e. it is $\propto b\times n_c$.
The number of clouds in each velocity bin sets the fluctuation
level in the spectrum. Thus, at a fixed level of fluctuations, for example
the one which corresponds to $\chi^2_r=1$, one expects $n_c\propto b^{-1}$.}
The ESI limits at $b<40$~\kms\ are nearly independent of $b$ because the
convolution with the instrumental line profile ($b=22.7$~\kms) leads to a nearly
constant line profile.
The upper limits on $n_c$ fall below the $n_c\propto b^{-1}$ relation at
$b>50$~\kms\ in all spectra due to the profile smoothing procedure, which filters out
large-scale profile structure. This effect is inevitable given the lack of 
{\it a priori}
knowledge about the intrinsic line profile. We note, however, that when $b>200$~\kms\
the simulated profile becomes distinctly different from the observed one. Our limits
are also subject to the assumption that all clouds produce similar line
profiles. Otherwise, any observed profile can be decomposed to any number of
different profile components, but such an arbitrary decomposition appears implausible.

We note in passing that the strongest constraint on $n_c$ (that is, the largest
contribution to $\chi^2_r$) comes from the observed fluctuations
in the line core, rather than the wings (cf. Dietrich et al. 1999; Ferland 2004).
This arises from the fact that the line flux density in the far wings
becomes smaller than the continuum flux density (here
$f^{\rm line}_{\lambda}>f^{\rm cont}_{\lambda}$
for $-900$~\kms$<v<600$~\kms\ in HIRES1, and for $-1100$~\kms$<v<1100$~\kms\ in HIRES2
and ESI). Thus, the far-wings noise is dominated by continuum photons
rather than line photons. Since the
continuum level is constant, the noise level is constant, and thus
the observed S/N is $\propto f_{\lambda}^{\rm line}$ (put in other terms,
the S/N in the far wings is background limited, rather than source limited).
In contrast, the MC simulation fluctuations
are anywhere proportional to the square root of the number of clouds contributing to
a given bin, and so are $\propto \sqrt{f_{\lambda}^{\rm line}}$. Thus, the ratio
of available S/N to the MC simulation fluctuations, which indicates the detectability
of the predicted fluctuations, is $\propto
\sqrt{f_{\lambda}}$; i.e. it approaches zero at the wings.

\section{Discussion}

The earlier studies of Arav et al. (1997, 1998) and Dietrich et al. (1999)
have already ruled out the bloated-stars scenario based on the lack of
the expected level of fluctuations in the line profiles. However, the fluctuation
level is reduced if the line width generated by each cloud is much larger
than the thermal width ($c_s\approx 10$\kms), leaving the possibility of a
clumped BLR if $\sigma\gg c_s$. Such a large $\sigma$ may be produced by, for
example, radiative ablation effects (e.g., Scoville \& Norman 1995; 
Pier \& Voit 1995).

The predicted level of fluctuations is $\propto 1/\sqrt{n_c}$. Since
$n_c \approx (R_{\rm BLR}/r_c)^2$, the largest fluctuations are expected in
objects with the smallest $R_{\rm BLR}$ --- that is, in the lowest-luminosity
objects. Here we report on analysis of Keck observations of the \ha\ profile
in NGC~4395, which is a factor of $\sim 10^3$ less luminous than the 
lowest-luminosity object previously analyzed (NGC~4151, Arav et al. 1998), 
and is the lowest-luminosity type~I AGN known by a significant factor
(compare with Barth et al. 2004; Greene \& Ho 2004).
Peterson et al. (2005) recently measured a size of
$1.0\times 10^{14}$~cm for the  $\mbox{C\,{\sc IV}}$ emitting region in
NGC~4395. The Balmer line emitting region is measured to be about
twice as large as the $\mbox{C\,{\sc IV}}$ emitting region in the few
objects were both lines were measured (Korista et al. 1995; Onken \& Peterson 2002),
and comparison of the \ha\ and $\mbox{C\,{\sc IV}}$ line widths suggests it could be as
much as 5 times larger here, so
one may expect $R_{\rm BLR} \approx (2-5)\times 10^{14}$~cm for the \ha\ line.
Thus, if the
bloated stars have $r_c \approx 10^{14}$~cm  (Alexander \& Netzer 1994), then
only a handful of them can fit in the BLR of NGC~4395, and this will produce
large profile features, essentially regardless of the line width generated by
each cloud. Such a model is clearly ruled out by the observed smoothness
of the \ha\ line, which implies $n_c\gtsim 10^4-10^5$ for $b \approx 10-100$~\kms.

The presence of $r_c \approx 10^{14}$~cm stars at such a small distance from the black hole
can also be ruled out based on tidal disruption arguments, as noted by Alexander \&
Netzer (1997). A star with a mass $M_*$ and radius $r_*$ will be tidally disrupted
once it crosses inward of $r_{\rm tidal} \approx r_*\times(M_{\rm BH}/M_*)^{1/3}$.
In NGC~4395, $M_{\rm BH}=3.6\times 10^5M_{\odot}$ (Peterson et al. 2005), implying
$r_{\rm tidal}/r_* \approx 70$,  and thus
bloated stars, which presumably have $M_* \approx M_{\odot}$\footnote{Significantly
more massive stars are unlikely due to their short lifetime, while compact stars
are unlikely due to their high surface gravity.}, will be disrupted already at
$r_{\rm tidal} \approx 7\times 10^{15}$~cm. Stars which survive down to  $R_{\rm BLR}$
without being tidally disrupted need to be smaller than $(3-7)\times 10^{12}$~cm,
or $(40-100)R_{\odot}$; they can at most be red giants, or blue supergiants.
A minimum population of $\sim 5000$ such stars is required in NGC~4395 to produce the
necessary BLR covering factor (using $R_{\rm BLR}/r_*\ge 70$). Such a population
will have a bolometric luminosity of $\sim 10^{39}-10^{40}$~erg~s$^{-1}$ (e.g.,
Table 3.14 of Binney \& Merrifield 1998), 
and its $I$-band magnitude will be comparable to (or larger
than) the unresolved point-source magnitude measured by Filippenko \& Ho (2003).
In any case, physical collisions will destroy such an extremely dense stellar system
($>10^{15}M_{\odot}$~pc$^{-3}$)  in a
few dynamical times (a few weeks). Thus, a stellar origin for the BLR clouds
in NGC~4395 is ruled out both on theoretical grounds, and directly through
the profile fluctuation analysis.

The tidal disruption argument can also be used to place a lower limit on the density of
any self-gravitating cloud at the BLR, regardless of its size. A cloud of density
$\rho_c$ and mass $M_c = 4\pi\rho_c r_c^3/3$ can avoid tidal disruption at the BLR if
$\rho_c>3/4\pi M_{\rm BH}/R_{\rm BLR}^3$. This gives a minimum
number density, $N_c>10^{18}$~cm$^{-3}$ in NGC~4395, about $10^8$ times larger than
the density indicated by photoionization models
(Kraemer et al. 1999). Thus, any clumped
emission-line gas in the BLR will be subject to the strong shearing effect of the
black hole gravity, which will stretch clouds by $\Delta r_c \approx r_c$ within one
dynamical time.

Our lower limit on $n_c$ of $\sim 10^4-10^5$ implies $r_c\ltsim (1-3)\times 10^{12}$~cm,
which is comparable to $d$, the minimum size implied by photoionization models.
Since we do not have a robust indication here that $r_c\ll d$, the BLR
gas may still be composed of clumps with the minimum acceptable size.
However, it is not clear what mechanism can
create, confine, and maintain such small individual clouds in a steady state
(e.g., Done \& Krolik 1996; Krolik 1999).
Thus, the most plausible scenario appears to be a smooth, or quasi-smooth,
gas flow. Reverberation mapping indicates that the BLR Balmer line velocity
is dominated by gravity, and a pure inflow
or outflow is excluded (e.g., Peterson 2001).
Thus, the only remaining velocity field appears to be a rotation-dominated
flow, most likely in a geometrically thick configuration
(to subtend the required covering factor). Direct evidence for an outflowing
warm ionized gas is present in UV (Crenshaw et al. 2004) and X-ray observations
(Shih et al. 2003), and this outflow may originate in the inner denser BLR
component.

A rotation-dominated gas flow should produce double-peaked lines.
Such lines are indeed observed in AGNs with very broad lines (e.g.,
Eracleous \& Halpern 2003; Strateva et al. 2003), but the majority of AGNs show
single-peaked lines. Chiang \& Murray (1996) suggested that a sufficiently
large radial velocity gradient can modify the line emissivity and
produce a single-peaked emission line. However, lower optical depth lines,
such as semi-forbidden lines and the higher H-series lines,
should still show double-peaked structure, unlike what is generally observed.

We note in passing that indirect indications for strongly blended ``hidden''
double peaks are provided
by the rotation of the polarization angle across the \ha\ line profile observed
in some objects (Smith et al. 2005), and by the distinct variability characteristics
of the red and blue line wings observed sometimes (e.g., Veilleux \& Zheng 1991;
Wanders \& Peterson 1996). Both effects were interpreted as evidence that
the Balmer lines originate in a rotating disk. The lack of a double-peak
structure could be induced by a large velocity broadening mechanism in the BLR,
which makes the double peaks blend into a single peak, although the double-peak 
structure would still be discernible through the different polarization and variability
characteristics of the two blended peaks. The two peaks appear to separate out in
objects where the Balmer lines are broad enough (e.g., Strateva et al. 2003).

To summarize, although
the BLR gas is likely to reside in a smooth flow, the origin of the gas,
the structure of the flow, and its dynamics remain to be clarified.

\acknowledgments
The data presented herein were obtained at the W. M. Keck Observatory, which is operated
as a scientific partnership among the California Institute of Technology, the University of
California, and the National Aeronautics and Space Administration. The Observatory was
made possible by the generous financial support of the W. M. Keck Foundation.
The authors wish to recognize and
acknowledge the very significant cultural role and reverence that the summit of Mauna Kea
has always had within the indigenous Hawaiian and the astronomical communities.
A.L. acknowledges support by the Israel Science Foundation
(Grant \#1030/04), and by a grant from
the Norman and Helen Asher Space Research Institute. A.V.F. is supported by NSF
grant AST-0307894; he is also grateful
for a Miller Research Professorship at U.C. Berkeley, during which
part of this work was completed.

\clearpage
\onecolumn
\begin{figure}
\includegraphics[angle=0,scale=.85]{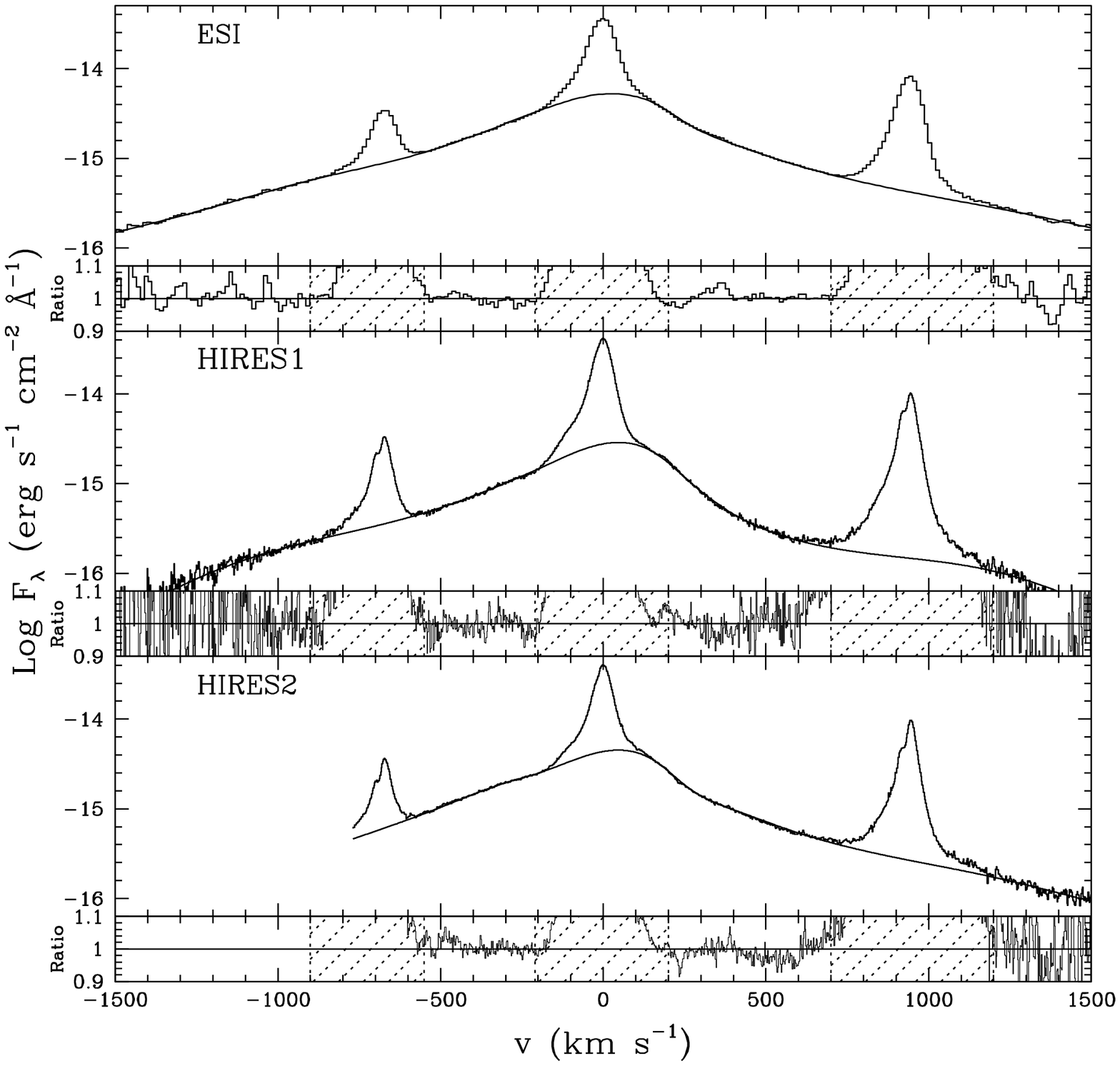}
\caption{The three Keck spectra of the \ha\ line used for the analysis.
The upper part of each subpanel shows the observed spectrum (histogram), and the
spline fit (smooth line). The lower part of each subpanel shows the
observed/spline-fit flux ratio; 
the hatched regions are affected by the
narrow emission lines and excluded from the fluctuation analysis.
Note that the ESI spectrum has the highest S/N ratio, although it has a lower
spectral resolution compared to the two HIRES spectra. The two HIRES spectra
are not flux calibrated; they were arbitrarily offset so that all 
spectra have a matching peak flux density.}
\end{figure}
\clearpage

\begin{figure}
\includegraphics[angle=270,scale=.65]{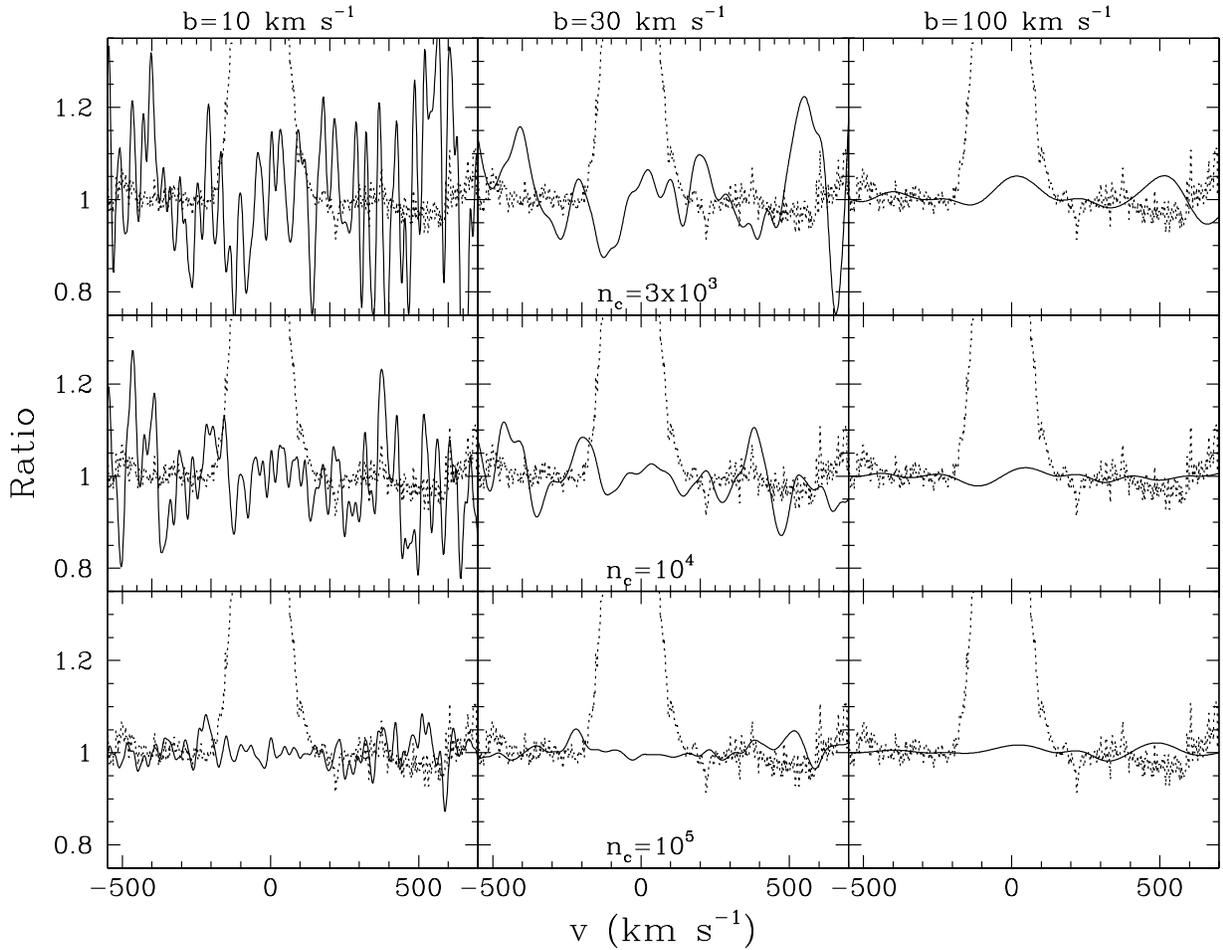}
\caption{A comparison of the observed versus simulated \ha\
profile fluctuations. The dotted lines show the observed/spline-fit flux ratio
for the HIRES2 spectrum. The solid lines shows the Monte Carlo 
simulation/spline-fit flux ratio.
Each row shows simulations with the same number of clouds $n_c$, and each
column shows simulations with the same Gaussian $b$ parameter. Note that at $b=10$~\kms\
(a pure thermal width), $n_c\gtsim 10^5$ is required to suppress the fluctuations
to the observed level; at $b=30$~\kms, $n_c>10^4$ is required; and at
$b=100$~\kms, $n_c\gtsim 3\times 10^3$ is sufficient.}
\end{figure}
 \clearpage

\begin{figure}
\includegraphics[angle=270,scale=.65]{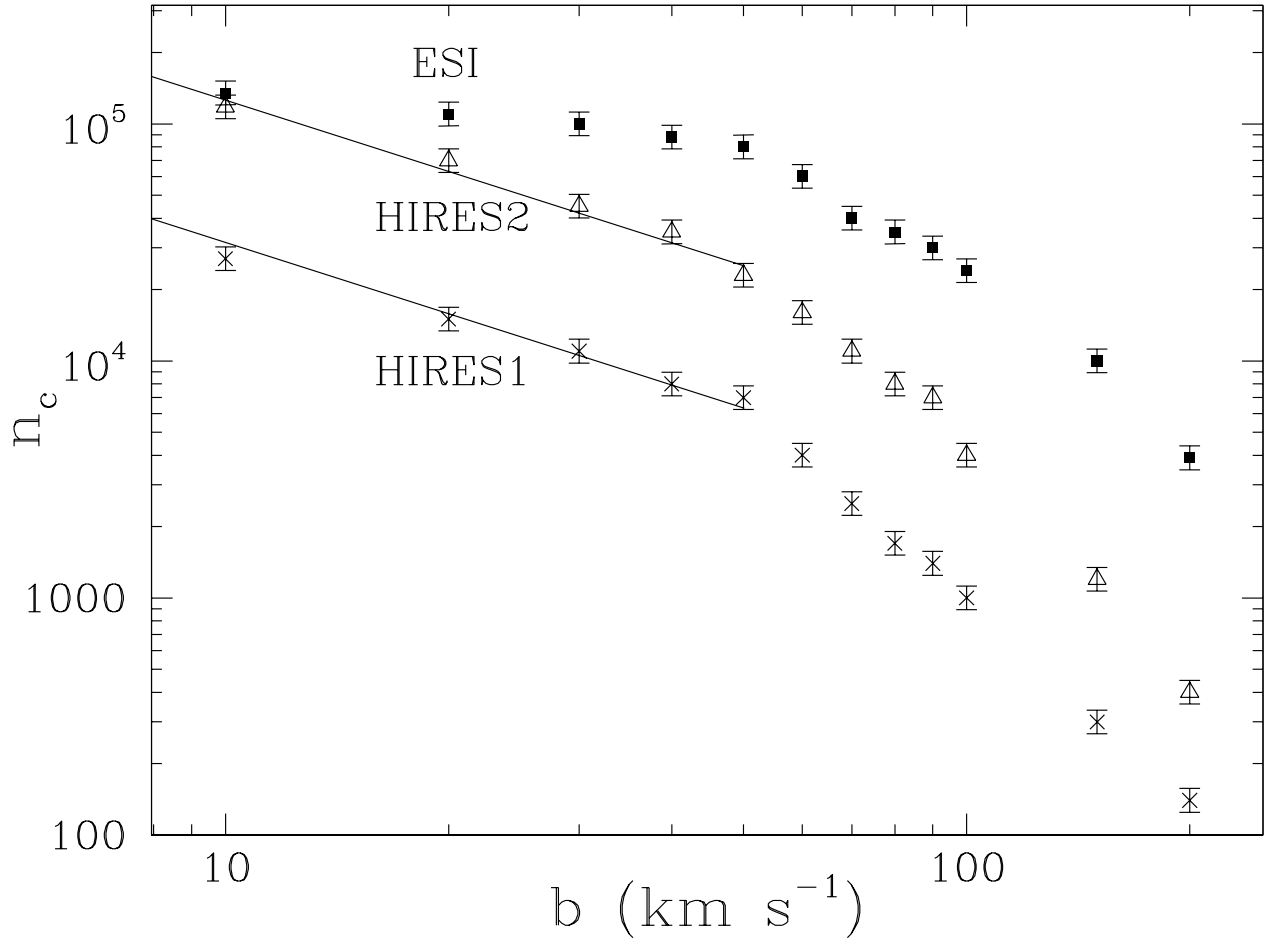}
\caption{The minimum number of clouds required to suppress the fluctuations
to the observed level, as a function of the width of the line emitted by
each cloud. The tightest constraints are obtained from the ESI spectrum, where
the broad \ha\ line has the highest S/N ratio (see Fig. 1). The solid lines follow
$n_c\propto b^{-1}$, the expected relation from simple fluctuation arguments.
The ESI limits at $b<40$~\kms\ are nearly independent of $b$ because the
convolution with the instrumental line profile leads to a nearly
constant effective $b$.
At $b>50$~\kms, the simulation constraints start falling below the $-1$
slope relation. This is a result of the spline smoothing which filters
out fluctuations on scales $\gtsim 100$~\kms.}
\end{figure}

\clearpage

\end{document}